\documentclass[aps,pra,twocolumn,footinbib]{revtex4-1}%
\usepackage{amsmath,amsfonts,amssymb}
\usepackage{graphicx,float,calc}
\usepackage{color,bm}
\usepackage[colorlinks,urlcolor=blue,citecolor=blue,linkcolor=blue]{hyperref}
\usepackage{amsmath}
\usepackage{amsfonts}
\usepackage{amssymb}
\usepackage{graphicx}
\usepackage{mathrsfs}
\usepackage[sort&compress]{natbib}%
\providecommand{\U}[1]{\protect\rule{.1in}{.1in}}
\begin{document}

	\author{Youjiang Xu }
\author{Han Pu}
\affiliation{Department of Physics and Astronomy, and Rice Center for Quantum Materials,
	Rice University, Houston, Texas 77251-1892, USA}

\title{Building Flat-Band Lattice Models from Gram Matrices}

\begin{abstract}
We propose a powerful and convenient method to systematically design flat-band
lattice models, which overcomes the difficulties underlying the previous
method. Especially, our method requires no elaborate calculations, applies to
arbitrary spatial dimensions, and guarantees to result in a completely flat
ground band. We use this method to generate several classes of lattice models,
including models with both short- and long-range hoppings, both topologically
trivial and non-trivial flat bands. Some of these models were previously
known. Our method, however, provides crucial new insights. For example, we have
reproduced and generalized the Kapit-Mueller model [Kapit and Mueller, Phys.
Rev. Lett. \textbf{105}, 215303 (2010)] and demonstrated a universal scaling
rule between the flat band degeneracy and the magnetic flux that was not
noticed in previous studies. We show that the flat band of this model results from the (over-)completeness properties of coherent states.

\end{abstract}
\maketitle

\section{Introduction}

Tight-binding lattice models that support flat bands
\cite{doi:10.1080/23746149.2018.1473052}, i.e., with single-particle energy
dispersion $E(\mathbf{k})$ independent of momentum $\mathbf{k}$, are of great
importance. The quenched kinetic energy and the associated macroscopic
degeneracy in a flat band makes the system extremely sensitive to
perturbations. In particular, in a many-body setting, interaction between
particles in the flat band, no matter how weak it is, can result in strong
correlations and exotic quantum phases. This is exactly what happens in, for
example, fractional quantum Hall systems
\cite{PhysRevLett.50.1395,PhysRevLett.63.199,RevModPhys.71.S298}, where the
underlying single-particle spectrum features flat Landau levels. It is
therefore important to understand what model Hamiltonians can support flat
bands and, conversely, how one can systematically design models supporting
flat bands.

At the first glance, it seems quite simple to construct a flat-band lattice
model in momentum space. Given dispersion relations $\left\{  E_{1}\left(
\mathbf{k}\right)  ,E_{2}\left(  \mathbf{k}\right)  ,...,E_{n}\left(
\mathbf{k}\right)  \right\}  $ in which one or a few of them are set to be
constant, e.g., $E_{1}\left(  \mathbf{k}\right)  =0$, we can construct a
$n$-band flat-band model from an arbitrary unitary matrix $U\left(
\mathbf{k}\right)  $: $h\left(  \mathbf{k}\right)  :=U\left(  \mathbf{k}%
\right)  E\left(  \mathbf{k}\right)  U^{\dag}\left(  \mathbf{k}\right)  $,
where $E\left(  \mathbf{k}\right)  $ is the diagonal matrix with elements
$\left\{  E_{1}\left(  \mathbf{k}\right)  ,E_{2}\left(  \mathbf{k}\right)
,...,E_{n}\left(  \mathbf{k}\right)  \right\}  $. However, this approach is
too general to be practical. Specifically, if we want the flat-band model to
possess certain properties (e.g., finite-range hopping, certain symmetry
properties, etc.), it is not straightforward to put appropriate constraints on
the matrices $U\left(  \mathbf{k}\right)  $ and $E\left(  \mathbf{k}\right)  $
to make $h\left(  \mathbf{k}\right)  $ possess the desired properties. To
solve this problem, we turn our attention from the momentum space to the real
position space.

Suppose the Hilbert space is spanned by the basis $\left\vert \mathbf{R}%
;i=1,2,\dots,n\right\rangle $, where $\mathbf{R}=\sum_{j=1}^{d}x_{j}%
\mathbf{e}_{j}$ ($x_{j}\in\mathbb{\mathbb{Z}}$) represents the position of
cells in a $d$-dimension lattice and $i$'s distinguish the $n$ sites in a
cell. In general, as long as a Hamiltonian $H$ defined on this basis satisfies
the translational symmetry $U_{\mathbf{R}}HU_{\mathbf{R}}^{\dag}=H$ for
arbitrary $\mathbf{R}$, where $U_{\mathbf{R}}\left\vert \mathbf{R}^{\prime
};i\right\rangle \equiv\left\vert \mathbf{R+R}^{\prime};i\right\rangle $, it
depicts a $n$-band system. Now it is not difficult to make the hopping terms
in $H$ finite ranged. This usually comes, however, at the cost of losing
control on the band dispersions.

Although it is difficult to manipulate the energy of all the $n$ bands
simultaneously in the position representation, it is still possible to set
some of the bands flat. It can be done by putting the bands into the null
space of the Hamiltonian. For example, the famous Lieb's lattice, whose flat
bands are crucial in proving magnetization when repulsive Hubbard interaction
is turned on \cite{PhysRevLett.62.1927.5}, consists of two sublattice $A$ and
$B$ with unequal numbers of sites and zero hopping magnitude between any two
sites in the same sublattice. This special bipartite connectivity results in
$\left\vert n_{A}-n_{B}\right\vert $ flat bands that reside in the null space
of the Hamiltonian, where $n_{A}$ and $n_{B}$ are the number of sites in the
two sublattice respectively. The flatness originates from the fact that the
Hamiltonian is a direct sum of $h_{AB}$ and $h_{BA}$, consisting of hopping
terms starting from sites in $A$ and $B$ respectively, and the rank of
$h_{AB}$ ($h_{BA}$) cannot be greater than the dimension of $h_{BA}$ ($h_{AB}%
$). Sutherland \cite{PhysRevB.34.5208} studies the band structure of general
bipartite systems and find that these connectivity-induced flat bands occur at
the middle of the spectrum, while there exist dispersive upper and lower bands
that are reflectively symmetric about the average on-site energy of the two
sublattices. It is remarkable that the flat bands of a bipartite lattice will
not be lifted by an external magnetic field because the flatness is not due to
fine-tuning of the hopping strengths \cite{PhysRevB.54.R17296}. However, a
general flat-band lattice model is not necessarily bipartite.

An alternative method to put a band into the null space of the Hamiltonian is
based on the following feature of any flat bands: One can superimpose Bloch
states on the flat band to make localized states which remain as eigenstates
on the same band. Such localized states that occupy fewest cells are called
compact localized states (CLSs). CLSs can be annihilated by their parent
Hamiltonian if the band energy is shifted to zero. Conversely, as long as we
find one of the CLSs, we can generate a whole family of them by translation,
each centering at a different cell. Altogether, this family of CLSs form a
flat band. Thus, using CLSs as generators, one can devise a systematic way of
constructing flat-band models. However, it is tricky to find appropriate CLSs.
For an arbitrary localized state, its parent Hamiltonian may not exist,
because the Hamiltonian is found by solving an inverse eigenvalue problem that
may not have a solution. Another drawback of the CLS method is that
information about the band spectrum cannot be obtained readily. In particular,
one cannot know \emph{a priori} whether the flat band is a ground or an
excited band. Moreover, the inverse problem is in general computationally
cumbersome, particularly for spatial dimensions larger than one. Only very
recently, the complete flat band generators on one dimensional (1D) lattice
was found \cite{PhysRevB.99.125129}, but its generalization to higher
dimensions is not straightforward. It is thus highly desirable to develop a
more powerful and convenient method of generating flat-band models.
\cite{PhysRevB.54.R17296,PhysRevB.95.115135,PhysRevB.99.125129,PhysRevLett.114.245503,Zong:16}%
.

In this paper, we will present such a new method based on the simple
mathematical properties of Gram matrices. The Gram matrix method put the
ground states into the null space of the Hamiltonian. It guarantees the
flatness of the lowest band by simple dimension-counting procedures without
invoking complicated inverse problems. Also, generating flat-band models in
high-dimensional lattices is straightforward as the method is insensitive to
spatial dimensions. The remaining of the paper is organized as follows. We
will describe the general principle of our method in Sec. II. In the next two
sections, we will present some specific examples to demonstrate its usage. In
Sec. III, we demonstrate the construction of several flat-band models with
finite-range hopping. In Sec. IV, we construct the long-range Kapit-Mueller
model in two-dimensional square lattice and present its generalization to
arbitrary lattice geometry. Finally, we conclude in Sec. V.

\section{General Principle of the Gram Matrix method}

A Gram matrix $G$ defines a semi-inner product in a linear vector space $V$.
Given a linear transformation $T:V\rightarrow V^{\prime}$, $G$ is the pullback
of the inner product defined in $V^{\prime}$, that is, ${G}\equiv{T}^{\dag}%
{T}$. Obviously, $G$ is Hermitian and semi-positive definite. Under a basis
$\{|v_{i}\rangle\}_{i=1}^{N}$ of the $N$-dimensional space $V$, the
$i^{\mathrm{th}}$ column of the $T$ matrix is the image of $|v_{i}\rangle$,
$\left\vert v_{i}^{\prime}\right\rangle :=T\left\vert v_{i}\right\rangle $, so
the matrix element of $G$ is the inner product $\langle v_{i}|G|v_{j}%
\rangle=\langle v_{i}^{\prime}|v_{j}^{\prime}\rangle$. If the set of vectors
$\{|v_{i}^{\prime}\rangle\}_{i=1}^{N}$ are linearly independent, then ${G}$ is
positive definite; otherwise, $G$ would be singular and possess zero
eigenvalues, and the number of zero eigenvalues equals the dimension of the
kernel of $T$. If $N^{\prime}<N$, where $N^{\prime}$ is the dimension of
$V^{\prime}$, then the set $\{|v_{i}^{\prime}\rangle\}_{i=1}^{N}$ is
necessarily linearly dependent, and the number of zero eigenvalues that ${G}$
possesses is at least $N-N^{\prime} $.

This simple property serves as the basic principle underlying our method. Up
to a shift of energy making its ground state energy zero, a Hamiltonian can
always be written as $H\equiv T^{\dag}T$ and can thus be interpreted as a Gram
matrix. Now $T$ is a linear transformation from the Hilbert space $V$ to an
auxiliary space $V^{\prime}$. We take the states $\left\vert \mathbf{R}%
;i=1,2,\dots,n\right\rangle $'s as the basis of $V$. If the auxiliary space
$V^{\prime}$ is spanned by sites $\left\vert \mathbf{R};i^{\prime}%
=1,2,\dots,n^{\prime}\right\rangle _{\text{aux}}$ in an auxiliary lattice,
where $n^{\prime}<n$, then we claim that \emph{any} $n$-band model with
$n-n^{\prime}$ flat lowest bands is associated with a matrix $T$ whose
elements are%
\begin{equation}
_{\text{aux}}\left\langle \mathbf{R}^{\prime};i^{\prime}\right\vert
T\left\vert \mathbf{R};i\right\rangle =T_{\mathbf{R-R}^{\prime}}^{i^{\prime
},i}\text{ .} \label{mapping}%
\end{equation}
Thus we can interpret the $T$ matrix graphically as hopping terms that
connects a single cell in the real lattice to sites in the auxiliary lattice.
Because $H\equiv T^{\dag}T$, as long as $T$ is finite-ranged, so will the
Hamiltonian $H$. We will construct some specific finite-range flat-band models
in Section III.

The proof our statement is straightforward: Any desired flat-band Hamiltonian
can be decomposed as $H=\sum_{\mathbf{k}}T_{\mathbf{k}}^{\dag}T_{\mathbf{k}}$,
where $T_{\mathbf{k}}$'s are $n^{\prime}$-by-$n$ matrices acting on
$\mathbf{k}$-Bloch states with the usual definition $\left\vert \mathbf{k}%
;i\right\rangle \equiv\sum\frac{e^{i\mathbf{k\cdot R}}}{\sqrt{N}}\left\vert
\mathbf{R};i\right\rangle $ in momentum space. Without loss of generality, we
can identify the image of $T_{\mathbf{k}}$ with the $\mathbf{k}$-Bloch states
in the auxiliary space $\left\vert \mathbf{k};i^{\prime}\right\rangle _{\text{
aux}}\equiv\sum\frac{e^{i\mathbf{k\cdot R}}}{\sqrt{N^{\prime}}}\left\vert
\mathbf{R};i^{\prime}\right\rangle _{\text{aux}}$. Thus $T=\sum_{\mathbf{k}%
}T_{\mathbf{k}}$ and the matrix elements of $T$ in Eq. (\ref{mapping}) are
related by a simple Fourier transformation. We can say that the translational
symmetry of the Hamiltonian is inherited by the $T $ matrix.

To summarize our Gram matrix method, here is the protocol to generate an
$n$-band model in $d$-dimensional space whose lowest few bands are degenerate:

\begin{enumerate}
\item The Hilbert space is spanned by $\left\vert \mathbf{R} ;i=1,2,\dots
,n\right\rangle $, where $\mathbf{R}$'s are positions of the unit cells and
each unit cell contains $n$ internal sites.

\item Construct an auxiliary space spanned by $\left\vert \mathbf{R};i^{\prime
}=1,2,\dots,n^{\prime}\right\rangle _{\text{aux}}$, where $n^{\prime}<n$.

\item A mapping $T$, whose matrix elements are given by Eq.~(\ref{mapping}),
is constructed to map between the Hilbert and the auxiliary spaces. Note that
$T$ is completely arbitrary unless some special properties are desired for the
resulting Hamiltonian.

\item The Hamiltonian can the constructed as $H=T^{\dag}T$. The lowest
$n-n^{\prime}$ bands of $H$ are guaranteed to be flat with energy 0.
\end{enumerate}

The above is the most general protocol to generate models with flat lowest
bands. The choice of the auxiliary space $V^{\prime}$ is flexible in the sense
that the space is not necessarily spanned by the lattice $\left\vert
\mathbf{R};i^{\prime}=1,2,\dots,n^{\prime}\right\rangle _{\text{aux}}$.
Especially, when the restriction on finite-range hopping is lifted, then an
alternative choice of $V^{\prime}$ may better serve our purpose. In Section
IV, we will see that if we select the auxiliary space to be spanned by a
subset of coherent states, then we can elegantly reproduce the Kapit-Mueller
model and its generalizations, revealing the nature of the massive degeneracy
in such models, which turns out to be a universal feature as required by the
(over-)completeness of the coherent states.

\section{Generating Finite-range Flat-band Models}

\begin{figure}[tbh]
\centering
\includegraphics[width=0.45\textwidth]{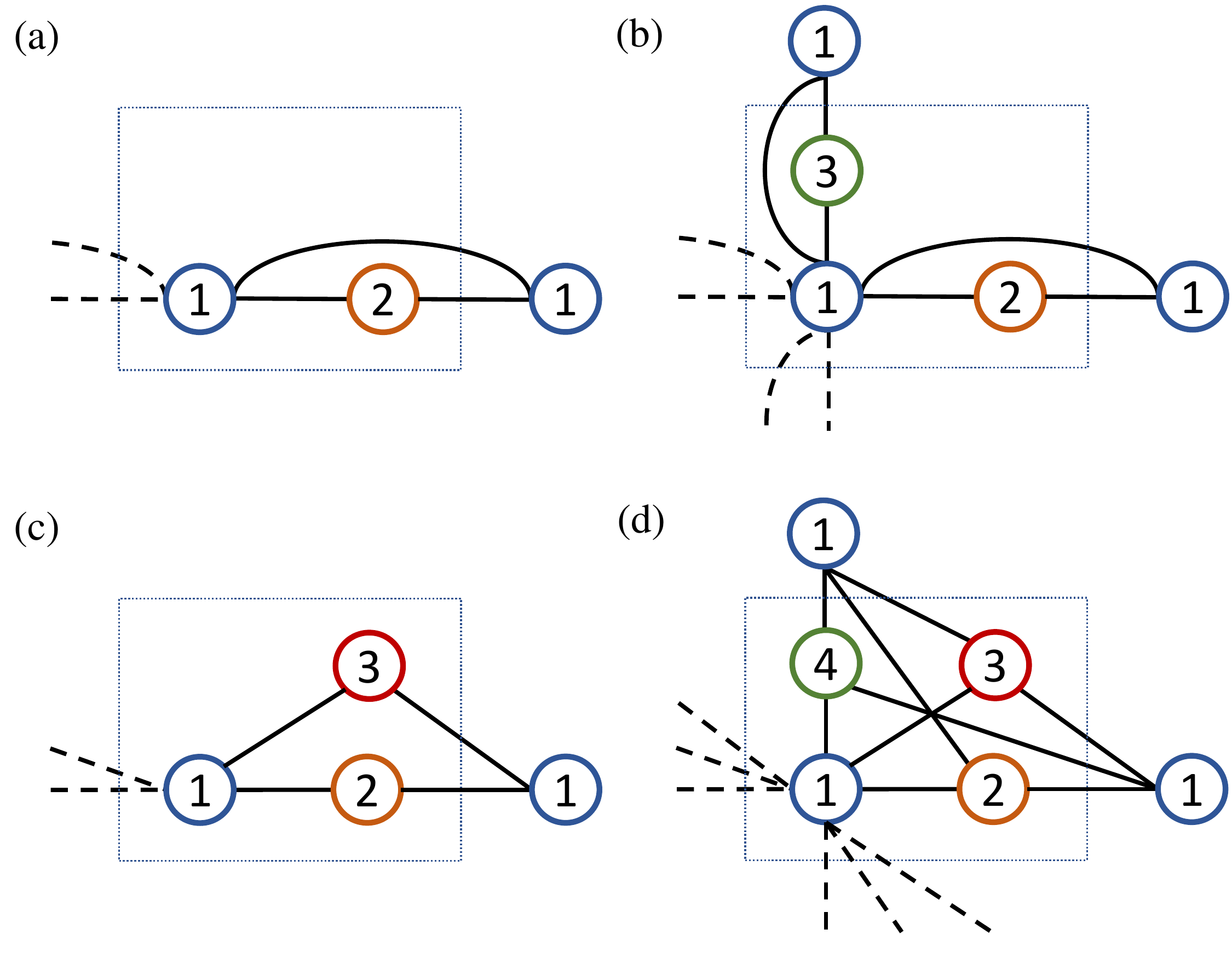}\caption{(Color online)
Tasaki lattice in 1D (a) and 2D (b) governed by Hamiltonian (\ref{tasaki}).
Bipartite lattice in 1D (c) and 2D (d) governed by Hamiltonian (\ref{bi}).
Rectangular boxes represent unit cells, with internal sites labelled by
circled numbers. Solid lines represent non-zero hoppings starting from a
single cell. Duplicated hopping terms are marked by dashed lines. }%
\label{lattice}%
\end{figure}

Let us now demonstrate the usage of the Gram matrix method by constructing
several specific models. In this section, we show the simplest models
constructed by the Gram matrix method. By simplest we mean that the model
Hamiltonian contains the fewest number of hopping terms in the $T$ matrix.
Considering a $d$-dimensional lattice, the simplest choice of the $T$ matrix
is $T\left\vert \mathbf{R};i\right\rangle =\left\vert \mathbf{R}%
;i\right\rangle _{\text{aux}}$. However, it is trivial because the resulting
Hamiltonian does not contain hopping terms between sites in the real lattice.
To obtain a non-trivial Hamiltonian, the $T$ matrix has to map the real
lattice cell at $\mathbf{R}$ to the auxiliary lattice cell at $\mathbf{R}$ and
to at least $d$ of its nearest neighbor cells. For example,
\begin{align*}
T\left\vert \mathbf{R};1\right\rangle  &  =\sum_{j=1}^{d}\left(
a_{j}\left\vert \mathbf{R};j\right\rangle _{\text{aux}}+b_{j}\left\vert
\mathbf{R+e}_{j};j\right\rangle _{\text{aux}}\right)  \text{,}\\
T\left\vert \mathbf{R};i\right\rangle  &  =\left\vert \mathbf{R}%
;i-1\right\rangle _{\text{aux}}\text{, }i=2,\dots,d+1\text{,}%
\end{align*}
where $a_{j},\,\,b_{j}\in\mathbb{\mathbb{C}}$. The corresponding Hamiltonian
is given by
\begin{align}
H &  =\sum_{\mathbf{R}}\left[  \sum_{i=1}^{d}\!\left(  \left\vert
a_{i}\right\vert ^{2}\!+\!\left\vert b_{i}^{2}\right\vert \right)  \left\vert
\mathbf{R};1\right\rangle \!\left\langle \mathbf{R};1\right\vert +\sum
_{i=2}^{d+1}\left\vert \mathbf{R};i\right\rangle \!\left\langle \mathbf{R}%
;i\right\vert \right]  \nonumber\\
&  +\sum_{\mathbf{R}}\sum_{i=1}^{d}\bigg{[}a_{i}\left\vert \mathbf{R}%
;i+1\right\rangle \left\langle \mathbf{R};1\right\vert \!+\!b_{i}\left\vert
\mathbf{R+e}_{i};i+1\right\rangle \left\langle \mathbf{R};1\right\vert
\nonumber\\
&  +a_{i}^{\ast}b_{i}\left\vert \mathbf{R+e}_{i};1\right\rangle \left\langle
\mathbf{R};1\right\vert \!+\!h.c.\bigg{]}.\label{tasaki}%
\end{align}
The underlying lattice is the $d$-dimensional Tasaki's lattice (examples in 1D
and 2D are presented in Fig.~\ref{lattice}(a) and (b), respectively), and the
hopping amplitudes in the original Tasaki's Hamiltonian
\cite{PhysRevLett.69.1608} represents a special case of Eq.~(\ref{tasaki})
with $a_{i}=b_{i}=1/\lambda$. This Hamiltonian has $n=d+1$ bands. Since each
unit cell in the auxiliary space has $n^{\prime}=d$ internal sites, $H$
possesses $n-n^{\prime}=1$ flat band. This flat band has zero energy and
represents the ground band of the system. The CLSs of the flat band can be
found as
\[
\left\vert \psi_{\mathbf{R}}^{0}\right\rangle \!=\!\left\vert \mathbf{R}%
;1\right\rangle -\!\sum_{i=1}^{d}\left(  a_{i}\left\vert \mathbf{R}%
;i\!+\!1\right\rangle +b_{i}\left\vert \mathbf{R+e}_{i};i\!+\!1\right\rangle
\right)  .
\]
If we group the sites $\left\vert \mathbf{R};1\right\rangle $ as sublattice
$A$ and the remaining sites $\left\vert \mathbf{R};i=2,3,...,d+1\right\rangle
$ as sublattice $B$, we find that the particle on a $B$ site can only hop to
an $A$ site. Following the same argument about the origin of the flat bands in
bipartite lattices, we conclude that there must be ($d-1$) additional flat
bands at energy one, whose CLSs are
\begin{align*}
\left\vert \psi_{\mathbf{R}}^{i}\right\rangle \! &  =\!a_{i+1}^{\ast
}\left\vert \mathbf{R};i\!+\!1\right\rangle +b_{i+1}^{\ast}\left\vert
\mathbf{R-e}_{i+1};i\!+\!1\right\rangle \\
&  -a_{i}^{\ast}\left\vert \mathbf{R};i\!+\!2\right\rangle -b_{i}^{\ast
}\left\vert \mathbf{R-e}_{i};i\!+\!2\right\rangle ,\;(i\!=\!1,...,d\!-\!1)
\end{align*}
Finally, there exists a dispersive top band with energy $E_{\mathbf{k}}%
=1+\sum_{i=1}^{d}\left\vert \alpha_{\mathbf{k}}^{i}\right\vert ^{2}$ where
$\alpha_{\mathbf{k}}^{i}:=a_{i}+b_{i}\exp\left(  -i\mathbf{k\cdot e}%
_{i}\right)  $. The corresponding eigenstate is
\[
\left\vert \psi_{\mathbf{k}}^{d}\right\rangle =\left(  E_{\mathbf{k}%
}-1\right)  \left\vert \mathbf{k};1\right\rangle +\sum_{i=1}^{d}%
\alpha_{\mathbf{k}}^{i}\left\vert \mathbf{k};i+1\right\rangle .
\]

The above example of constructing the $d$-dimensional Tasaki's model serves a
pedagogical purpose, from which we see the power of the Gram matrix method
that it takes no extra effort to generalize the model to high dimensions,
whereas the conventional CLS method involves solving more and more complicated
equation when $d$ increases. By modifying the $T$ matrix, any model with a
desired number of lowest few flat bands can be readily constructed. We can
also artificially put constraints on the $T$ matrix so that the resulting
model possesses desired properties. To this end, we now turn to construct a
finite-range hopping lattice model which possesses a special feature: all of
its bands are flat.

The basic idea rooted in the reflection symmetry of the upper and lower bands
in a bipartite model. From \cite{PhysRevB.34.5208} we know the middle bands in
a bipartite model are flat, and if there is only one lower band (i.e., the
ground band) whose flatness is guaranteed by the Gram matrix, then the only
upper band must also be flat because of the reflection symmetry. So in order
to build a model in which all bands are flat, we are going to design a
bipartite structure which guarantees $n-2$ middle flat bands and, by choosing
$n^{\prime}=n-1$, there is a zero-energy flat ground band. Such a $T $ matrix
can be chosen as follows:%
\begin{align*}
T\left\vert \mathbf{R};1\right\rangle  &  =c_{d+1}\left\vert \mathbf{R}%
;d+1\right\rangle _{\text{aux}}+\sum_{j=1}^{d}c_{j}\left\vert \mathbf{R+e}%
_{j};j\right\rangle _{\text{aux}}\text{,}\\
T\left\vert \mathbf{R};i\right\rangle  &  =\sum_{j=1}^{d+1}u_{i-1,j}\left\vert
\mathbf{R};j\right\rangle _{\text{aux}}\text{, }i=2,3,\dots,d+2\text{,}%
\end{align*}
where $c_{j}\in\mathbb{\mathbb{C}}$ and $u_{i,j}$ is an arbitrary unitary
matrix. The unitary matrix guarantees the bipartite structure: Different rows
of $u_{i,j} $ are orthogonal to each other, so the hopping amplitudes between
sites $\left\vert \mathbf{R};i=2,3,\dots,d+2\right\rangle $ are zero, and the
onsite energy of these sites is uniform because each row of the unitary matrix
has the same norm. The resulting Hamiltonian is given by:
\begin{align}
H  &  =\sum_{\mathbf{R}}\sum_{i=1}^{d+1}\left\vert c_{i}\right\vert
^{2}\left\vert \mathbf{R};1\right\rangle \left\langle \mathbf{R};1\right\vert
+\sum_{\mathbf{R}}\sum_{i=2}^{d+2}\left\vert \mathbf{R};i\right\rangle
\left\langle \mathbf{R};i\right\vert \nonumber\\
&  +\sum_{\mathbf{R}}\sum_{i=1}^{d+1}\bigg{[}u_{i,d+1}^{\ast}c_{d+1}\left\vert
\mathbf{R};i+1\right\rangle \left\langle \mathbf{R};1\right\vert \nonumber\\
&  +\sum_{j=1}^{d}u_{i,j}^{\ast}c_{j}\left\vert \mathbf{R+e}_{j}%
;i+1\right\rangle \left\langle \mathbf{R};1\right\vert +h.c.\bigg{]}\text{ ,}
\label{bi}%
\end{align}
The lattice connectivity in 1D and 2D are illustrated in Fig.~\ref{lattice}(c)
and (d), respectively. The bipartite nature can be easily seen if we group
$\left\vert \mathbf{R};1\right\rangle $ as sublattice $A$ with onsite energy
$E_{A}=\sum_{i}^{d+1}\left\vert c_{i}\right\vert ^{2}$ , and $\left\vert
\mathbf{R};i=2,..,d+2\right\rangle $ as sublattice $B$ with onsite energy
$E_{B}=1$. This $\left(  d+2\right)  $-band model consists of a flat ground
band with energy zero, $d$ flat bands with energy $E_{B}=1$, and a flat top
band with energy $E_{A}+E_{B}=1+\sum_{i}^{d+1}\left\vert c_{i}\right\vert
^{2}$, reflecting the reflection symmetry of bipartite lattices.

\section{The Kapit-Mueller Model and its Generalization}

We have seen two examples of finite-range flat-band models. The flat bands in
a finite-range Hamiltonian are necessarily non-topological with zero Chern
number. This is due to a theorem \cite{Chen_2014} which states that the
following three conditions concerning a band cannot be simultaneously
satisfied: (1) Being flat; (2) Having a non-zero Chern number; (3) The
Hamiltonian is finite-range. To create a flat band with finite Chern number,
it is then necessary to construct a model with infinite-range hopping
amplitudes. In general, as we have demonstrated in the Introduction, a flat
topological ground band can always be obtained by writing down a topologically
non-trivial unitary matrix $U\left(  \mathbf{k}\right)  $ to construct the
Hamiltonian in the momentum space: $h\left(  \mathbf{k}\right)  =U\left(
\mathbf{k}\right)  E\left(  \mathbf{k}\right)  U^{\dag}\left(  \mathbf{k}%
\right)  $, where $E\left(  \mathbf{k}\right)  =$ diag$\left(  0,E_{2}\left(
\mathbf{k}\right)  ,E_{3}\left(  \mathbf{k}\right)  ,\dots,E_{n}\left(
\mathbf{k}\right)  \right)  $. Usually, we want the constructed Hamiltonian to
be short ranged, in the sense that the hopping strength decays fast when the
hopping distance increases. It is, however, inconvenient to work in the
momentum space to restrain the hopping distance.

In \cite{PhysRevLett.105.215303}, Kapit and Mueller, working in the real
space, found such a topological flat band in a 2D square lattice, and
attributed the massive degeneracies in the flat band to some unrevealed
symmetries. It was realized that the degenerate ground states can be regarded
as discrete lowest Landau levels (LLLs), and that the degeneracy of the LLLs
give birth to the flatness. Here we find an alternative way to understand the
origin of this topological flat band by reproducing the model with Gram
matrices. More specifically, we reproduce the Kapit-Mueller model by a Gram
matrix built upon a subset of coherent states. From this construction, the
massive degeneracy of the ground band can be straightforwardly understood as a
result of the linear dependency of the coherent states. And we find the
generalizations of the model, which are beyond the LLL descriptions, can also
share the ground band degeneracy.

\subsection{Constructing Kapit-Mueller model on arbitrary 2D lattice}

A coherent state $\left\vert z\right\rangle $ is an eigenstate of a bosonic
annihilation operator with complex eigenvalue $z$. It is well known that the
full set of coherent states form an overcomplete basis. Perelomov
\cite{Perelomov1971} studied the completeness of a countable subset of
coherent states. Define
\[
z_{m,n}:=m\omega_{1}+n\omega_{2},\;\;m,n\in\mathbb{\mathbb{Z}},\;\;\omega
_{1},\omega_{2}\in\mathbb{\mathbb{C}}.
\]
$z_{m,n}$'s form a 2D lattice on the complex plane whose unit cell area is
$S:=\mathrm{Im}\,\omega_{1}^{\ast}\omega_{2}$. We collect the set of coherent
states $\{\left\vert z_{m,n}\right\rangle \}$. Perelomov find that: If
$S\leq\pi$, the set represents an overcomplete basis; If $S>\pi$, the set is
incomplete; If $S=\pi$, we can take away any one of the $\left\vert
z_{m,n}\right\rangle $'s from the set, and the remaining states form a
complete basis.

Following the basic procedures described in Section II, and replacing the
orthonormal basis of the auxiliary space by the lattice of coherent states, we
consider a linear transformation $T$ that maps $\left\vert m,n\right\rangle
:=\left\vert m\mathbf{e}_{1}+n\mathbf{e}_{2};1\right\rangle $, a site on a 2D
lattice containing single state in a cell, to $\left\vert z_{m,n}\right\rangle
$'s:
\begin{equation}
T\left\vert m,n\right\rangle =\left\vert z_{m,n}\right\rangle \,. \label{Tkm}%
\end{equation}
In other words, the columns of the $T$ matrix are formed by $\left\vert
z_{m,n}\right\rangle $'s. The matrix elements of the Hamiltonian $H=T^{\dag}T$
are therefore the inner product of coherent states:
\begin{align}
\left\langle m^{\prime},n^{\prime}\right\vert H\left\vert m,n\right\rangle  &
=\left\langle z_{m^{\prime},n^{\prime}}|z_{m,n}\right\rangle \nonumber\\
&  =e^{-\left\vert z_{m,n}-z_{m^{\prime},n^{\prime}}\right\vert ^{2}%
/2+i\mathrm{Im}\,z_{m^{\prime},n^{\prime}}^{\ast}z_{m,n}}\text{.} \label{H_KM}%
\end{align}
Physically, this Hamiltonian describes a fully connected 2D lattice under a
magnetic field, and the flux per unit cell is $2S$. It reduces to the
Kapit-Mueller model \cite{PhysRevLett.105.215303} when $z_{m,n}$'s form a
square lattice, and to the Hofstadter model \cite{PhysRevB.14.2239} by further
taking the limit $S\rightarrow\infty$.

From our construction, it immediately becomes clear that the emergence of the
flat ground band is guaranteed by the properties of the Gram matrix and the
(over-)completeness of the coherent states. When $S>\pi$, the set of coherent
states on lattice are linearly independent, and the smallest eigenvalue of the
resulting $H$ must be positive. When $S=\pi$, the set becomes complete if we
take away any one of the states, so $H$ has a single zero eigenvalue. When
$S<\pi$, we have $1/S$ states per unit area on the complex plane, while only
$1/\pi$ states per unit area are needed to construct a complete basis, so a
fraction of $\rho\equiv1-S/\pi$ eigenvalues of $H$ must be zero. As a result,
the massive degeneracy in the Hamiltonian Eq. (\ref{H_KM}) follows a universal
scaling behavior, in the sense that it only depends on $S$ and is completely
independent of the lattice geometry. As we will see, it is more convenient to
interpret $S$, instead of as the unit cell area, as the averaged area on the
complex plane occupied by each coherent state, because the relation between
$S$ and the completeness extends beyond the cases that Perelomov studied. The
universal degeneracy is one of the most elegant features of this model.

\begin{figure}[tbh]
\centering
\includegraphics[width=0.4\textwidth]{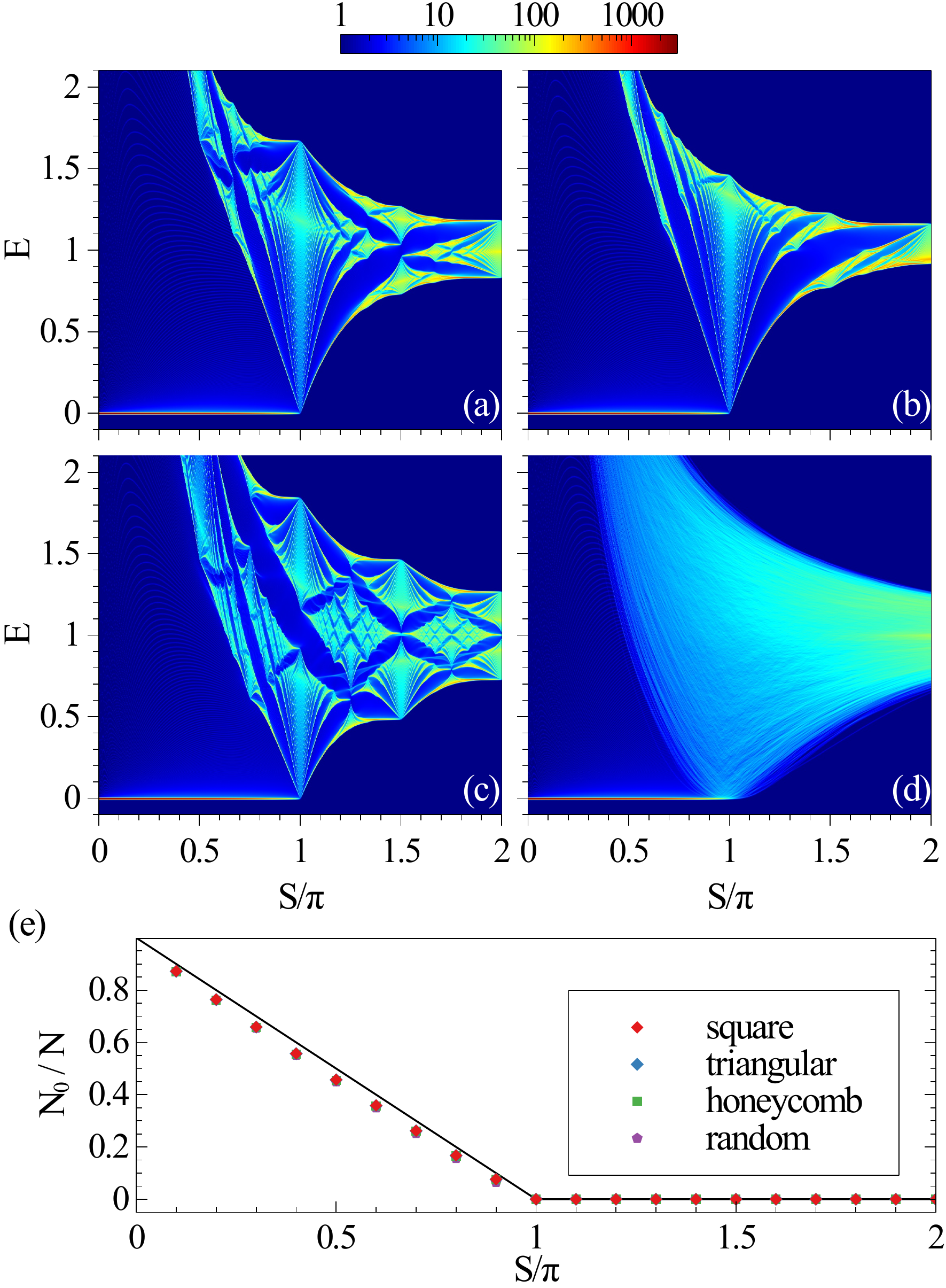}\caption{(Color online)
(a)-(d) Spectrum of $H^{S}$ for a Square lattice (a), Triangular lattice (b),
Honeycomb lattice (c), and Random lattice (d). The color map represents
$(D+1)$ where $D(E)=\sum_{n}\delta(E-E_{n}) $ is the density of states, with
$E_{n}$ being the $n^{\mathrm{th}} $ eigenvalue of $H^{S}$. In the
calculation, the Dirac $\delta$-function is replaced by a smooth narrow
distribution function. $S$ is the averaged area per state. (e) We count
$N_{0}$, the number of eigenenergy that is less than $10^{-5}$, and compare
the ratio $N_{0}/N$ with the theoretical value $\rho$ marked by the line,
where $\rho=\max(1-S/\pi,0)$. }%
\label{spectrum}%
\end{figure}

We carry out numerical calculation to verify the universal scaling rule
between the degeneracy and $S$. We choose $N$ coherent states distributed in a
square region on the complex plane, where $N$ is large but finite (typically,
$N\sim3\times10^{3}$), and numerically diagonalize the corresponding Gram
matrix $H$ to find the density of states as a function of $S$ and the energy
$E$. In Fig.~\ref{spectrum}(a)-(c), we display the spectrum for several
distinct lattice geometry (The lattice geometry refers to the geometry of the
coherent states on the complex plane.): square lattice, triangular lattice,
and honeycomb lattice. The positive-energy part of the spectrum forms a
Hofstadter butterfly, whose specific pattern depends on the lattice geometry.
The universal feature for these different lattices is, however, the massively
degenerate ground states at zero energy when $S<\pi$. Remarkably, this flat
ground band exists even when the coherent states has a random distribution
over the whole region (we specify a lower bound on the distances between sites
to ensure that no two sites are too close together in order to exclude trivial
zero eigenvalues of the Gram matrix) as we show in Fig.~\ref{spectrum}(d).
Although in this random lattice case, the positive-energy butterfly pattern no
longer exists.

As mentioned earlier, the degeneracy of the ground band is given by
$N_{0}=\rho N=(1-S/\pi)N$, which should be a universal feature independent of
the lattice geometry. In Fig.~\ref{spectrum}(e), we plot the numerically
obtained fraction of zero-energy states as a function for $S$. Results from
all four lattice geometries show excellent agreement with the theoretical
prediction. The small discrepancies can be attributed to the finite-size
effect. In Fig.~\ref{spectrum}(a)-(d), we see a gap between the zero-energy
and the positive-energy states, which increases as $S$ decreases and diverges
when $S\rightarrow0$. This can be easily understood as follows: Since
$\mathrm{Tr}\left[  H\right]  =N$ which is the sum of the positive
eigenenergies, the averaged energy of the excited states should be $\pi/S$
when $S<\pi$ according to the degeneracy of the ground states.

We also carry out the same numerical calculation on a small $5\times5$ square
lattice and see that the degenerate ground states emerge again. This
demonstrates that the essential physics can be observed even in such a small
system, which makes its experimental realization very promising.

\subsection{Universal ground states and the Hall dynamics}

The analogue between the zero-energy ground states of the model and the LLLs
can be found by writing down the wave functions explicitly. Using a more
general singlet sum rule \cite{LAUGHLIN1989163} proved by
Perelomov~\cite{Perelomov1971}:
\[
\sum_{m,n}\left(  -1\right)  ^{m+n+mn}e^{-\frac{\left\vert \alpha
_{m,n}\right\vert ^{2}}{2}+\alpha_{m,n}z}\equiv0\,,
\]
where $z$ is an arbitrary complex number and $\alpha_{m,n}$'s build an
arbitrary lattice whose unit cell area is $\pi$, we can show that the wave
function of the CLS in the ground band takes the universal form ($S<\pi$):
\begin{equation}
\left\langle m,n|\psi\right\rangle =\left(  -1\right)  ^{m+n+mn}%
\,e^{-\frac{\pi/S-1}{2}\left\vert z_{m,n}\right\vert ^{2}}\,.\label{lll1}%
\end{equation}
Ignoring the phase factor $\left(  -1\right)  ^{m+n+mn}$, Eq.~(\ref{lll1})
takes the same form as the wave function of a LLL describing a particle with
unit charge confined in the $\left(  x,y\right)  $-plane subjected to a
perpendicular magnetic field with strength $B$:
\begin{equation}
\psi_{\mathrm{LLL}}\left(  x,y\right)  =e^{-B\left(  x^{2}+y^{2}\right)
/4\hbar}\,.\label{lll2}%
\end{equation}
Also, just like the LLL, the ground band formed by $\left\vert \psi
\right\rangle $ and its translations is topological with Chern number 1. If
the nondimensionalization is done by setting $2\pi\hbar$, and the unit cell
area of the lattice formed by $\left\vert m,n\right\rangle $'s \footnote{The
distance between physical sites is not necessarily the same as the distance
between $z_{m,n}$'s on the complex plane. Here we make it independent from
$S$.} as $1$, then by comparing the exponents in Eqs.~(\ref{lll1}) and
(\ref{lll2}), we find that the degeneracy of the flat band per site $\rho$ can
be regarded as the effective magnetic field for $\left\vert \psi\right\rangle
$, which is different from the true flux $2S$. Moreover, it is this effective
field $\rho$, rather than the true flux, that plays a role in the Hall
dynamics of $\left\vert \psi\right\rangle $. Given a constant electric field
$\mathscr{E}$, the Hall velocity of $\left\vert \psi\right\rangle $ is
$\mathscr{E}/\rho$, because the Hall conductivity is $1$. Similarly, the Hall
velocity for $\psi_{\mathrm{LLL}}$ is $\mathscr{E}/B$.

\begin{figure}[tbh]
\centering
\includegraphics[width=0.4\textwidth]{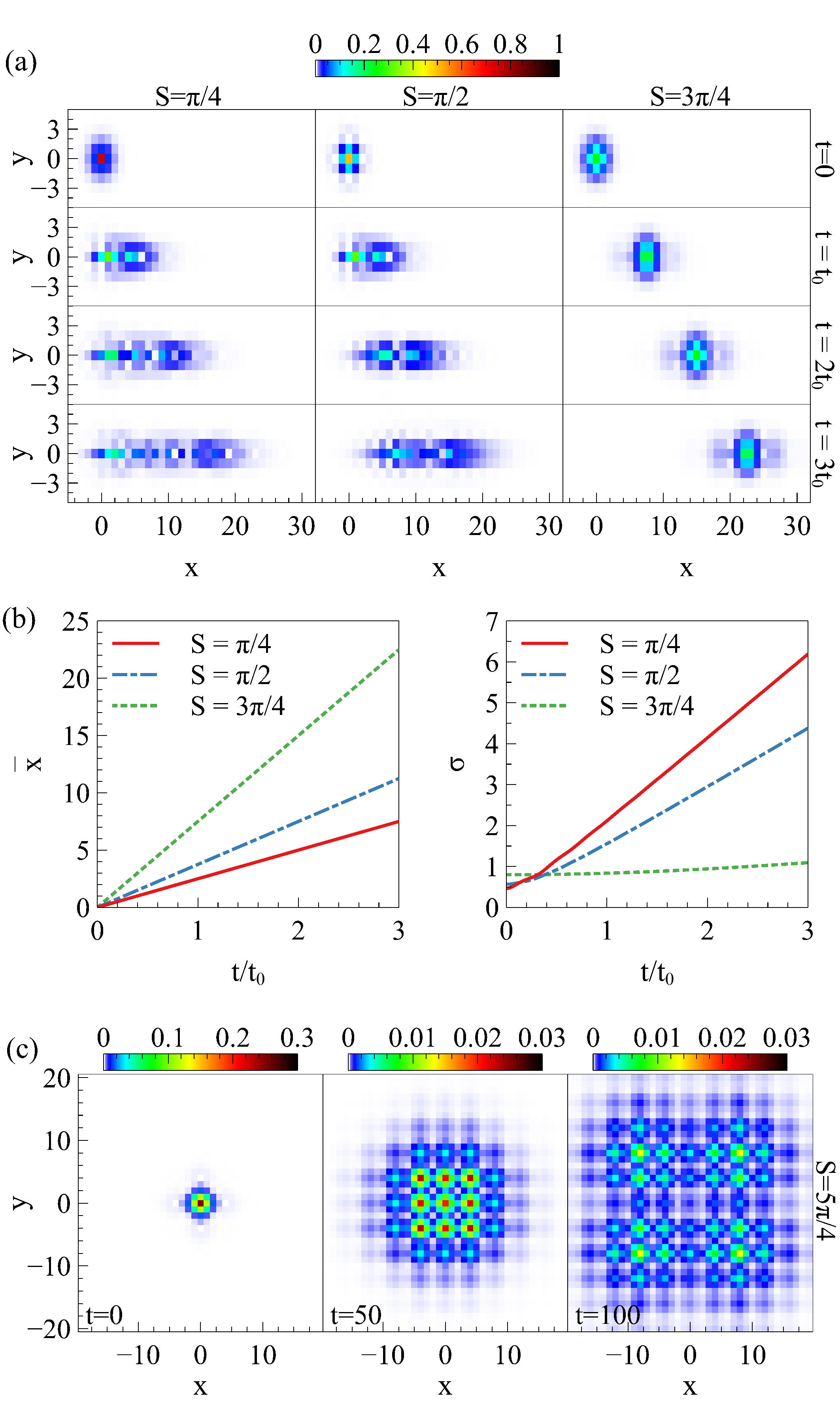}\caption{(Color online) (a)
Evolution of the density profile of a wavepacket on an $L\times L=81\times81$
square lattices with a linear potential along the $y$-axis with gradient $U$.
The evolution operator is given by $\exp\{-2\pi it(G^{S}+Uy)\}$, where we take
$U=3.75\times10^{-4}$. The positions $x$ and $y$ are renormalized by the
lattice constant. The initial wavepacket is obtained by projecting a
completely localized state to the lowest (zero-energy) band. $t_{0}=5000$. (b)
The evolution of the central position $\bar{x}$ and the half width $\sigma$
along the $x$ direction of the wavepacket in (a). (c) The same evolution as
that in (a) except that now $S=5\pi/4$ such that the lowest band is not a flat
band.}%
\label{Hall}%
\end{figure}

The Hall dynamics of $|\psi\rangle$ can be verified numerically as followes.
We initially prepare a wavepacket in the zero-energy ground band localized at
the center of an $81\times81$ square lattice. We then add a linear potential,
with gradient $U$, along the $y$-axis at $t=0$, and the ensuing evolution of
the wavepacket is depicted in Fig.~\ref{Hall}. Fig.~\ref{Hall}(a) show the
snapshots of density profiles at various times. The three columns correspond
to $S=\pi/4$, $\pi/2$ and $3\pi/4$ from left to right. One can see that the
wavepacket disperses and moves along the $x$-axis, perpendicular to the
direction of the linear potential. The evolution of the center position
$\bar{x}$ along the $x$-axis is plotted in the left panel of Fig.~\ref{Hall}%
(b). $\bar{x}$ is nearly linear in time, which results from the very accurate
quantization of the Hall conductivity. We can readily extract the Hall
velocities, $v_{H}=5\times10^{-4},\,7.5\times10^{-4},\,1.5\times10^{-3}$
respectively. The Hall conductivity, given by $v_{H}(1-S/\pi)/U$, is equal to
1 with a relative standard deviation on the order $10^{-7}$ for all three
cases. As the wavepacket moves, it also disperses along the $x$-axis (but not
along the $y$-axis). The right panel of Fig.~\ref{Hall}(b) displays the
evolution of the half width $\sigma$ along the $x$-axis. The slower the
wavepacket moves, the faster it disperses. Note that the LLL wavepacket is
non-dispersive under a similar situation. In Fig.~\ref{Hall}(c), we also
display the evolution of a localized wavepacket for $S=5\pi/4$. In this case,
the wavepacket quickly disperses in both directions, in stark contrast against
the situation depicted in Fig.~\ref{Hall}(a) where $S<\pi$.

\subsection{Analytic derivation of the full spectrum for special cases}

Despite the numerical evidence that supports universal degeneracy presented in
the generalized Kapit-Mueller model, it is helpful if we can prove the
degeneracy and write down the Bloch wave functions by solving the Hamiltonian
Eq.~(\ref{H_KM}) directly. It can be done for some special choices of
$\left\vert z_{m,n}\right\rangle $'s.

Consider the case when $\rho$ is rational by taking $S=p\pi/q$ where $p$, $q$
are co-prime positive integers. After a local gauge transformation $\left\vert
m,n\right\rangle \rightarrow e^{-imnS}\left\vert m,n\right\rangle $ and then a
Fourier transformation, $H$ can be reduced to a $q$-band Bloch Hamiltonian
$h^{p,q}\left(  k,l\right)  $, where $k$, $l$ are pseudo momenta defined in
the range $-\frac{1}{q}<k\leq\frac{1}{q}$, $-1<l\leq1$. The explicit matrix
elements of $h^{p,q}$ are given by:
\begin{align}
h_{m^{\prime},m}^{p,q}\!  &  =\!\sum_{r,s}\exp
\!\big{\{}\!-\!|z_{qr+m-m^{\prime},s}|^{2}/2+i\pi\big{[}k(qr+m-m^{\prime
})\nonumber\\
&  +(l+(m+m^{\prime})pq)s-prs\big{]}\big{\}},\,(m,m^{\prime}%
=1,2,...,q)\nonumber
\end{align}

If $z_{m,n}$'s form a rectangular lattice with aspect ratio $\xi$ on the
complex plane, then $h_{m^{\prime},m}^{p,q}$ can be expressed through Jacobi
$\theta$-functions (See the Appendix for definitions):
\begin{align*}
h_{m^{\prime},m}^{p,q}  &  =\exp\left(  -\frac{S\xi\left(  m-m^{\prime
}\right)  ^{2}}{2}+i\pi k\left(  m-m^{\prime}\right)  \right) \\
&  \times\sum_{\chi=0,1}\theta_{3+\chi\left(  p\operatorname{mod}2\right)
}\left(  z_{1},\tau_{1}\right)  \theta_{3-\chi}\left(  z_{2},\tau_{2}\right)
\end{align*}
where%
\begin{align*}
\tau_{1}  &  :=\frac{ipq\xi}{2}\,\,,\qquad z_{1}:=\frac{qk}{2}+\frac{\left(
m-m^{\prime}\right)  \tau_{1}}{q}\,,\\
\qquad\tau_{2}  &  :=\frac{2ip}{q\xi},\qquad z_{2}:=l+\frac{\left(
m+m^{\prime}\right)  p}{q}.
\end{align*}

We are going to verify that when $p=1$, $h_{m^{\prime},m}^{p,q}$ is a rank-$1$
matrix, so it has only one non-zero eigenvalue, which is predicted by the Gram
matrix construction. Now $p=1$, $\tau_{1}=-\tau_{2}^{-1}$, so we can use
Jacobi identities%
\[
\theta_{3+\chi}\left(  \frac{z}{\tau},-\frac{1}{\tau}\right)  =\left(
-i\tau\right)  ^{1/2}\exp\left(  i\pi z^{2}/\tau\right)  \theta_{3-\chi
}\left(  z,\tau\right)  \text{, }%
\]
where $\chi=0,\pm1$, to recast $h^{1,q}$ as follows%
\begin{align*}
h_{m^{\prime},m}^{1,q}\! &  =\!\left(  -i\tau_{2}\right)  ^{1/2}\sum
_{\chi=0,1}\theta_{3-\chi}\left(  z_{1}\tau_{2}\right)  \theta_{3-\chi}\left(
z_{2}\right)  \\
&  \times\exp\!\left(  \!-\frac{S\xi\!\left(  m-m^{\prime}\right)  ^{2}}%
{2}+i\pi k\left(  m-m^{\prime}\right)  +i\pi z_{1}^{2}\tau_{2}\right)  ,
\end{align*}
where, for simplicity, we have omitted the second argument $\tau_{2}$ of the
$\theta$-functions. Define
\begin{align*}
R_{m,m^{\prime}} &  :=\frac{i}{\tau_{2}}\left(  h_{m^{\prime},m}%
^{1,q}h_{m^{\prime}+1,m+1}^{1,q}-h_{m^{\prime}+1,m}^{1,q}h_{m^{\prime}%
,m+1}^{1,q}\right)  \\
&  \times\exp\left(  S\xi\left(  m-m^{\prime}\right)  ^{2}-2i\pi\left(
k\left(  m-m^{\prime}\right)  +z_{1}^{2}\tau_{2}\right)  \right)  .
\end{align*}
$R_{m,m^{\prime}}\equiv0$ iff $h^{1,q}$ is rank-1. Representing $h^{1,q}$ by
the $\theta$-functions, we have
\begin{align}
\!\!R_{m,m^{\prime}}\!\!=\!\!\sum_{\chi,\chi^{\prime}} &  \theta_{3-\chi
}\!\left(  z_{1}\tau_{2}\right)  \theta_{3-\chi}\!\left(  z_{2}\right)
\theta_{3-\chi^{\prime}}\!\left(  z_{1}\tau_{2}\right)  \theta_{3-\chi
^{\prime}}\!\left(  z_{2}+\frac{2}{q}\right)  \nonumber\\
&  -\theta_{3-\chi}\left(  z_{1}\tau_{2}+\frac{1}{q}\right)  \theta_{3-\chi
}\left(  z_{2}+\frac{1}{q}\right)  \nonumber\\
&  \times\theta_{3-\chi^{\prime}}\left(  z_{1}\tau_{2}-\frac{1}{q}\right)
\theta_{3-\chi^{\prime}}\left(  z_{2}+\frac{1}{q}\right)  .\label{sumrule1}%
\end{align}
Using the sum rules of the $\theta$-functions, presented in the Appendix, we
can prove:
\[
R_{m,m^{\prime}}=0.
\]
So we have proved that, indeed, for rectangular lattice with aspect ratio
$\xi$, $h^{1,q}$ has only one positive eigenvalue, whose value is given by the
trace:
\[
\mathrm{Tr}\left(  h^{1,q}\right)  =q\sum_{r,s}\exp\left(  -\frac{1}%
{2}\left\vert z_{qr,qs}\right\vert ^{2}+i\pi q\left(  kr+ls-rs\right)
\right)  .
\]
The corresponding Bloch wave function is given by any column vector of
$h^{1,q}$.

\subsection{Generalization beyond the LLL analogy}

The Kapit-Mueller model can also be constructed using a projection method
\cite{PhysRevA.88.033612} which is analogous to the inverse method by treating
the LLLs as CLSs. However, using the Gram matrix method, infinite variations
of the model, which cannot be regarded as the parent Hamiltonians of the LLLs,
can be constructed a similar protocol described in Sec.~II. In the discussion
above to construct the Kapit-Mueller model, we have picked a particular $T$
matrix defined in Eq.~(\ref{Tkm}). However, we can define the $T$ matrix,
which maps the real cell $\left\vert m,n\right\rangle $ to an arbitrary number
of cells represented by the coherent states in the auxiliary space, in a more
general way as follows:
\begin{equation}
T\left\vert m,n\right\rangle =\sum_{m^{\prime},n^{\prime}}\tau_{m^{\prime
},n^{\prime}}^{m,n}\left\vert z_{m^{\prime},n^{\prime}}\right\rangle \text{.}%
\label{newT}
\end{equation}
The resulting Hamiltonian is $H'=\tau^{\dag}H\tau$, where $H$ is the Gram matrix
of the coherent states given in Eq.~(\ref{H_KM}) which yields the Kapit-Mueller model. Apparently, the new
Hamiltonian $H'$ has at least the same number of zero eigenvalues as that of $H$.
In order that the zero-energy states of $H'$ form a flat band, $H'$
must preserve the translational symmetry of $H$, which contrains the $T$ matrix in Eq.~(\ref{newT}). The translational symmetry is
preserved as long as:%
\begin{equation}
\Phi_{\mathbf{R}}^{\dag}U_{\mathbf{R}}^{\dag}\tau U_{\mathbf{R}}%
\Phi_{\mathbf{R}}=\tau, \label{translation}%
\end{equation}
where $U_{\mathbf{R}}$ is the translation operator and $\Phi_{\mathbf{R}}$
a local gauge transformation as the result of the gauge flux through the
lattice:%
\[
\Phi_{p\mathbf{e}_{1}+q\mathbf{e}_{2}}\left\vert m,n\right\rangle =\exp\left(
-iS\left(  np-qm\right)  \right)  \left\vert m,n\right\rangle ,
\]
where $\mathbf{e}_{1,2}$ are the two basis vectors of the lattice, and $p, q \in \mathbb{\mathbb{Z}}$.
To find $\tau$'s that preserve the translational symmetry, we recast $\tau$ into the form%
\begin{equation}
\tau_{m^{\prime},n^{\prime}}^{m,n}=\sum_{r,s}C\left(  m,n;r,s\right)
\delta_{m^{\prime}}^{m+r}\delta_{n^{\prime}}^{n+s}\text{.}%
\end{equation}
Now we have \begin{widetext}
\[
\left.  \left(  \Phi_{\mathbf{R}}^{\dag}U_{\mathbf{R}}^{\dag}\tau
U_{\mathbf{R}}\Phi_{\mathbf{R}}\right)  _{m^{\prime},n^{\prime}}%
^{m,n}\right\vert _{\mathbf{R=}p\mathbf{e}_{1}+q\mathbf{e}_{2}}=\sum
_{r,s}C\left(  m+p,n+q;r,s\right)  \exp\left(  iS\left(  sp-qr\right)
\right)  \delta_{m^{\prime}}^{m+r}\delta_{n^{\prime}}^{n+s}\text{.}%
\]
\end{widetext}
To make Eq. (\ref{translation}) hold, $C\left(  m,n;r,s\right)  $ must take the
form:%
\begin{equation}
C\left(  m,n;r,s\right)  = C\left(  r,s\right) \, \exp\left[ iS\left(
rn-sm\right)  \right]  \text{,}%
\end{equation}
where $C\left(  r,s\right)  $ is an arbitrary function.

\begin{figure}[tbh]
\centering
\includegraphics[width=0.28\textwidth]{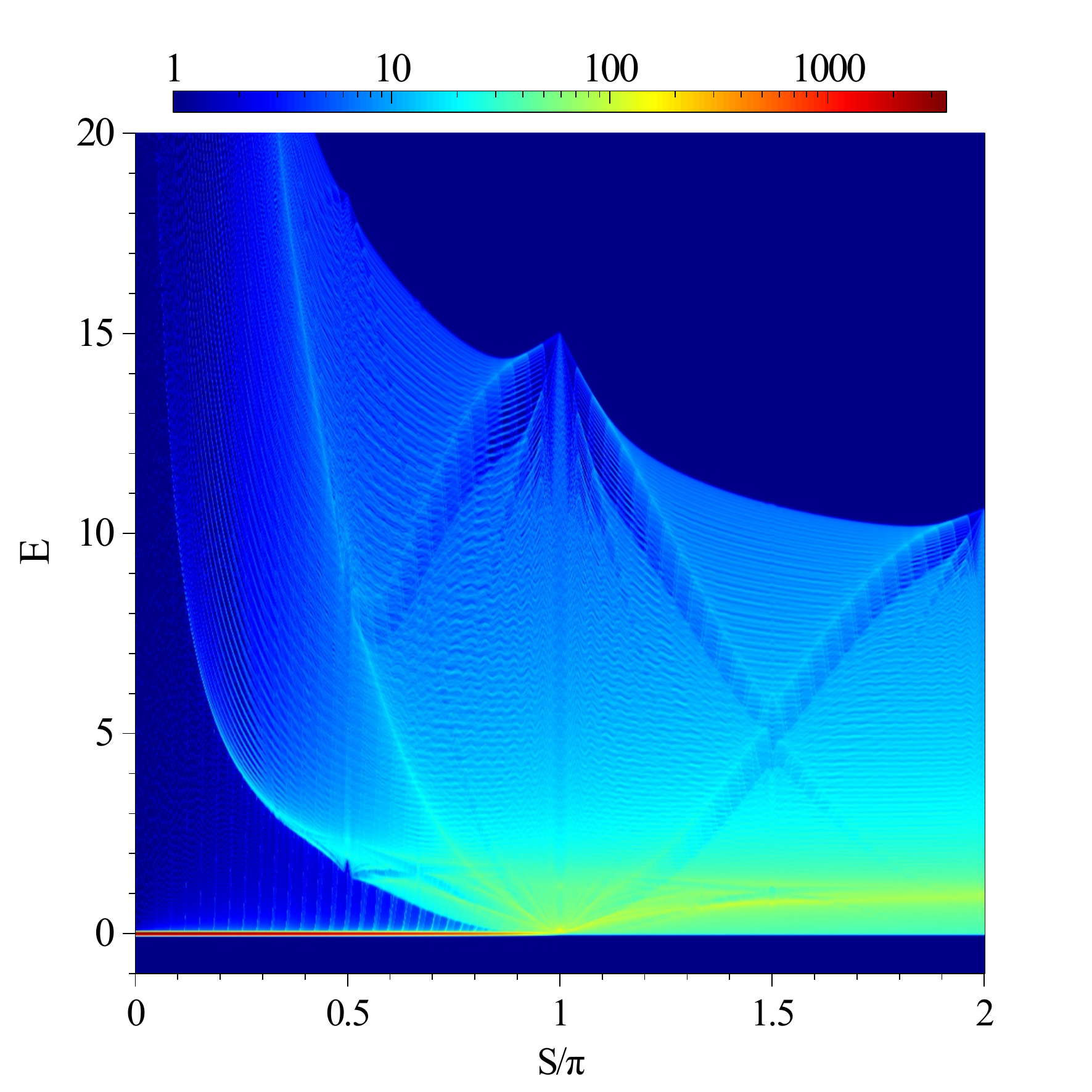}\caption{(Color online) The
spectrum of the generalized Kapit-Mueller model given by the $T$ matrix Eq.~(\ref{generalized T}). The color map has the same meaning as that in Fig.~\ref{spectrum} (a). }
\label{generalized spectrum}%
\end{figure}

For example, analogous to the 2D Tasaki's lattice constructed in Sec. III,
a $T$ matrix that maps a real cell $\left\vert m,n\right\rangle $ to the
auxiliary cell $\left\vert z_{m,n}\right\rangle $ and its two nearest
neighbors is given by
\begin{equation}
\label{generalized T}T\left\vert m,n\right\rangle =\left\vert z_{m,n}%
\right\rangle + e^{ iSn} \left\vert z_{m+1,n}\right\rangle
+e^{ -iSm } \left\vert z_{m,n+1}\right\rangle .
\end{equation}
The resulting Hamiltonian's zero-energy band has the same degeneracy as that of the
Kapit-Mueller model. In Fig.~\ref{generalized spectrum}, we illustrate the spectrum of this Hamiltonian on a square lattice by plotting the density of states as a function of the unit cell area $S$.

\section{Conclusion}

We have proposed a powerful method of constructing lattice models supporting
flat bands. The method is based on the mathematical properties of Gram
matrices. Any lattice model with flat lowest band can be constructed through
the method. The method does not require any elaborate calculations such as
solving the inverse eigenvalue problem, works for arbitrary spatial
dimensions, and guarantees to produce a flat ground band. We have presented a
variety of examples, including both finite- and infinite-range hopping,
topologically trivial and nontrivial flat bands. Specifically, we have
constructed the $d$-dim Tasaki lattice, a $d$-dim bipartite lattice whose
bands are all flat, and the generalized Kapit-Mueller lattice whose flat
ground band features universal (i.e., geometry-independent) degeneracy. We
study the generalized Kapit-Mueller in detail and, especially, conclude that
the (over-)completeness of the coherent states is the origin of the universal degeneracy.

Finally, we want to comment on realizing flat-band models in laboratory. Over
the past few years, we have witnessed rapid progress in realizing lattice
models in synthetic materials, particularly synthetic dimensions, in both
atomic
\cite{PhysRevLett.112.043001,Mancini1510,Stuhl1514,PhysRevLett.115.095302,Barbarino_2016,PhysRevA.99.013624,sundar2018synthetic,PhysRevLett.120.040407,Lohse2018,Sugawa1429}
and photonic
\cite{lustig2019photonic,PhysRevA.87.013814,luo2015quantum,Yuan16,PhysRevA.93.043827,Yuan18,PhysRevLett.118.083603,PhysRevA.95.062120,PhysRevA.100.043817,PhysRevX.6.041043,Schine2016}
systems, where the lattice sites are represented by different atomic states or
photonic modes, respectively. Nearly arbitrary hopping amplitudes can be
realized in such systems. Realizing the flat band models constructed using our
method with synthetic materials should therefore pose no essential difficulties.

We acknowledge the support from the NSF and the Welch Foundation (Grant No. C-1669).

%

\appendix*

\section{Jacobi $\theta$-function}

\begin{widetext}
The Jacobi $\theta $-functions are defined by%
\begin{eqnarray*}
\theta _{1}\left( z,\tau \right) &:&=\sum_{n}\exp \left( \pi i\tau \left(
n+1/2\right) ^{2}+2i\pi \left( z-\frac{1}{2}\right) \left( n+1/2\right)
\right) \\
\theta _{2}\left( z,\tau \right) &:&=\sum_{n}\exp \left( \pi i\tau \left(
n+1/2\right) ^{2}+2i\pi z\left( n+1/2\right) \right) \\
\theta _{3}\left( z,\tau \right) &:&=\sum_{n}\exp \left( \pi i\tau
n^{2}+2i\pi zn\right) \\
\theta _{4}\left( z,\tau \right) &:&=\sum_{n}\exp \left( \pi i\tau
n^{2}+2i\pi \left( z-\frac{1}{2}\right) n\right)
\end{eqnarray*}%
\end{widetext}The sum rules of the $\theta$-functions used in the paper are%
\begin{align*}
&  \theta_{2}\left(  v+w\right)  \theta_{2}\left(  v-w\right)  \theta
_{2}\left(  x+y\right)  \theta_{2}\left(  x-y\right) \\
&  -\theta_{2}\left(  x+w\right)  \theta_{2}\left(  x-w\right)  \theta
_{2}\left(  v+y\right)  \theta_{2}\left(  v-y\right) \\
&  =\theta_{1}\left(  v+x\right)  \theta_{1}\left(  v-x\right)  \theta
_{1}\left(  y+w\right)  \theta_{1}\left(  y-w\right)
\end{align*}%
\begin{align*}
&  \theta_{3}\left(  v+w\right)  \theta_{3}\left(  v-w\right)  \theta
_{3}\left(  x+y\right)  \theta_{3}\left(  x-y\right) \\
&  -\theta_{3}\left(  x+w\right)  \theta_{3}\left(  x-w\right)  \theta
_{3}\left(  v+y\right)  \theta_{3}\left(  v-y\right) \\
&  =-\theta_{1}\left(  v+x\right)  \theta_{1}\left(  v-x\right)  \theta
_{1}\left(  y+w\right)  \theta_{1}\left(  y-w\right)
\end{align*}%
\begin{align*}
&  \theta_{2}\left(  v+w\right)  \theta_{3}\left(  v-w\right)  \theta
_{3}\left(  x+y\right)  \theta_{2}\left(  x-y\right) \\
&  -\theta_{2}\left(  x+w\right)  \theta_{3}\left(  x-w\right)  \theta
_{3}\left(  v+y\right)  \theta_{2}\left(  v-y\right) \\
&  =-\theta_{4}\left(  v+x\right)  \theta_{1}\left(  v-x\right)  \theta
_{1}\left(  y+w\right)  \theta_{4}\left(  y-w\right)
\end{align*}
By taking
\begin{align*}
v  &  =z_{1}\tau_{2}\,,\qquad w=0\,,\qquad\\
x  &  =z_{2}+\frac{1}{q}\,,\qquad y=\frac{1}{q}\text{, }%
\end{align*}
one can readily show that $R_{m,m^{\prime}}$ in Eq.~(\ref{sumrule1}) indeed vanishes.
\end{document}